\documentclass[5p]{elsarticle} 

\usepackage{graphics} 
\usepackage{amsmath} 
\usepackage{amssymb}  
\usepackage{color}
\usepackage{fancyhdr}

\definecolor{dkgreen}{rgb}{0,0.6,0}
\definecolor{gray}{rgb}{0.5,0.5,0.5}
\definecolor{mauve}{rgb}{0.58,0,0.82}

\renewcommand{\citet}[1]{\citeauthor{#1} (\citeyear{#1})}
\renewcommand{\citep}[1]{(\citeauthor{#1} \citeyear{#1})}

\newcommand{\nvec}{\boldsymbol{n}}
\newcommand{\xvec}{\boldsymbol{x}}

\begin{document}

\begin{frontmatter}




\title{Probabilistic Cross-Identification of Galaxies with Realistic Clustering 
}


\author[math,cs]{Neil Mallinar}
\author[ams,cs,pha]{Tam\'{a}s Budav\'{a}ri}
\author[pha]{Gerard Lemson}

\address[math]{Dept.~of Mathematics, Johns Hopkins University, USA }
\address[ams]{Dept.~of Applied Mathematics \& Statistics, Johns Hopkins University, USA}
\address[cs]{Dept.~of Computer Science, Johns Hopkins University, USA}
\address[pha]{Dept.~of Physics \& Astronomy, Johns Hopkins University, USA}

\begin{abstract}
Probabilistic cross-identification has been successfully applied to a number of problems in astronomy from matching simple point sources to associating stars with unknown proper motions and even radio observations with realistic morphology. 
Here we study the Bayes factor for clustered objects and focus in particular on galaxies to assess the effect of typical angular correlations. 
Numerical calculations provide the modified relationship, which (as expected) suppresses the evidence for the associations at the shortest separations where the 2-point auto-correlation function is large. Ultimately this means that the matching probability drops at somewhat shorter scales than in previous models.
\end{abstract}


\end{frontmatter}

\section{Motivation}
In the new age of surveys, the identification of sources across telescopes and instruments is one of the most fundamental problems in observational astronomy. Most time-domain and multicolor studies rely on high-quality associations of catalogs with trustworthy and intuitive quality measures, e.g., probability.
The flexible framework introduced by \citet{BS08-BayesCrossID} uses Bayesian hypothesis testing to find the most likely associations. An analytic formula can be derived in the usual limits of astrometric models, which enables efficient execution that takes the same time as previous (sometimes adhoc) matching methods. Simulations have shown that the Bayesian approach handles variable uncertainties well \citep{heinis} 
and other studies illustrated its power on scenarios that could not be addressed before: matching stars without knowing their proper motions \citep{kerekes}, incorporating photometry of galaxies \citep{marquez} and even radio morphology \citep{fan}; also see \citet{budavari_loredo} for a review. 

The imaging data provide a likelihood function for each detection $\ell_i(\nvec)$ where the argument is the (unknown) true celestial position in the sky. This is frequently summarized by the direction of the detection $\xvec_i$ (weighted average of pixel directions) and an astrometric uncertainty $\sigma_i$ assuming a Gaussian, hence \mbox{$\ell_i(\nvec)\!=\!G(\xvec_i;\nvec,\sigma_i^2)$}. 
In the flat-sky approximation, we can deal with the more general case of elliptical errors, when the likelihood function $\ell_i(\nvec)\equiv{}p(\xvec_i|\nvec) = G(\xvec_i; \nvec, \Sigma_i)$ depends on the covariance matrix, and hence
\begin{equation}
\ell_i(\nvec) = \frac{1}{\sqrt{\lvert{}2\pi\Sigma_i\rvert}}\,\exp\left\{ -\frac{1}{2} (\xvec_i\!-\!\nvec)^T \Sigma_i^{-1} (\xvec_i\!-\!\nvec)\right\} \,.
\label{eq:multivariate-gaussian}
\end{equation}
In the spherical case, potentially warranted by large uncertainties, one can use the \citet{fisher} distribution; see \citet{BS08-BayesCrossID}. The following discussions apply to these situations the same way.

Following \citet{BS08-BayesCrossID}, we consider two hypotheses and compare the marginal likelihoods by forming their ratio, the so-called Bayes factor.
The numerator is the likelihood of the hypothesis that the detections belong to the same source in direction $\nvec$, and the denominator is that of the alternative, which claims that they are two separate objects at $\nvec_1$ and $\nvec_2$. 
Assuming identical sky coverage of the observations and a corresponding $p(\nvec)$ prior, this factor is
\begin{equation}
B = \frac{ \int p(\nvec)\ \ell_1(\nvec)\,\ell_2(\nvec)\ d\nvec } 
{ \int\!\!\int p(\nvec_1)\,p(\nvec_2)\ \ell_1(\nvec_1)\,\ell_2(\nvec_2)\ d\nvec_1\,d\nvec_2 } \,.
\label{eqn:expanded_original_bayes}
\end{equation}
%
When $B$ is large, the data favor the match, and the posterior probability is larger than the prior; see discussion in \citet{BS08-BayesCrossID}.
If $B$ ratio is less than one, the no-match hypothesis is preferred in the sense that the posterior is even lower than the prior.
The above formula is, however, not completely general as the aforementioned studies have primarily focused on a special case in the limit where the objects' locations are assumed to be independent.
This simplified the calculations and an analytic solution was possible for Gaussian and Fisher likelihood functions. 

Here we study the effect of spatial clustering on the Bayes factor and the posterior probability, when the independent assumption is replaced by an angular 2-point correlation function.
In Section~\ref{sec:evidence} we discuss the assumptions and derive the formalism. Section~\ref{sec:eval} provides an overview of how our new model was evaluated, and Section~\ref{sec:sum} discusses the effect on the probabilities and matching procedures.

\section{Galaxy Clustering in Matching}
\label{sec:evidence}

Galaxies are clustered in space, which is reflected also in their angular distribution on the sky. 
We incorporate this effect into the Bayes factor calculation using a realistic power-law model for the angular two-point correlation function.

We start by defining the general directional Bayes factor
\begin{equation}
B' = \frac{\int p(\nvec)\ \ell_1(\nvec)\,\ell_2(\nvec)\,d\nvec}
{\int\!\!\int p(\nvec_1,\nvec_2)\ \ell_1(\nvec_1)\,\ell_2(\nvec_2)\ d\nvec_1\,d\nvec_2}, 
\label{eq:bf_new}
\end{equation}
where the parameters of the alternative hypothesis $\nvec_1$ and $\nvec_2$ can depend on each other via the join prior density function, which is written as the product
\begin{equation}
p(\nvec_1,\nvec_2)=p(\nvec_1)\,p(\nvec_2|\nvec_1) 
\end{equation}
by definition.
The conditional prior $p(\nvec_2 | \nvec_1)$ accounts for the fact that galaxies are clustered. 
Of course, the new integral is typically not separable and simple analytic calculations are not possible. First we look at how the physical model enters the conditional $p(\nvec_2|\nvec_1)$ prior and numerically evaluate the new integrals to illustrate the effects of galaxy clustering.

The 2-point angular correlation function $w(\theta)$ represents the excess probability of finding a galaxy on the sky at $\theta$ radians away from another galaxy \citep{peebles}. With that, the conditional prior probability density can be written as
\begin{equation} \label{eq:prior}
p(\nvec_2 | \nvec_1) = \frac{1}{4\pi}\Big[1 + w(\theta_{12})\Big]
\end{equation}
where $\theta_{12}$ is the angular separation between unit vectors $\nvec_1$ and $\nvec_2$, i.e., \mbox{$\nvec_1\!\cdot\!\nvec_2\!=\!\cos\theta_{12}$}. 
It is important to note that this is a hypothetical correlation function of the model prior and not a measurement. The measurement problems associated with the correlation functions are outside of the scope of this study. Here we simply assume a standard parametrization to assess the importance of the clustering.

We adopt the usual power-law parametrization to the correlation function 
\begin{equation}
w{(\theta)} = A\,\left({\frac{\theta}{\theta_0}}\right)^{\!\!-\delta} \!- C \,,
\label{eqn:correlation_power_law}
\end{equation}
where $C$ is a constant due to the \emph{integral constraint}
\begin{equation}
\int w(\theta)\sin\theta\ d\theta = A \ \theta_0^\delta \ I_\delta - 2 C = 0 \,,
\end{equation}
which is derived from the fact that the conditional prior in eq.~(\ref{eq:prior}) integrates to 1, and $w(\theta)$ to 0, over the sphere. Here
\begin{equation}
I_\delta \equiv \int\limits_{0}^{\pi} \theta^{-\delta} \sin\theta\ d\theta, 
\end{equation}
so
\begin{equation}
C = \frac{1}{2} A \ \theta_0^{\delta} \ I_\delta \,.
\end{equation}
This \emph{integral constraint} can be calculated for any $(A, \delta)$ parameters given a constant $\theta_0$ value used for normalization.
%
%

For our prior distribution, we take the fiducial parameters from the Sloan Digital Sky Survey \citep{york00} 
study by \citet{2003ApJ...595...59B} and fix \mbox{$\delta\!=\!0.8$} and \mbox{$A\!=\!0.08$} for the purpose of this analysis, using a constant value \mbox{$\theta_0\!=\!0.1^{\circ}$}.  
This results in 
a constant of \mbox{$C\!\approx\!4.4\!\cdot\!10^{-4}$}, which is indeed very small and proved to be negligible in our analysis.

\begin{figure*}
\centering
\includegraphics[trim=20mm 0mm 20mm 7mm,clip,width=\textwidth]{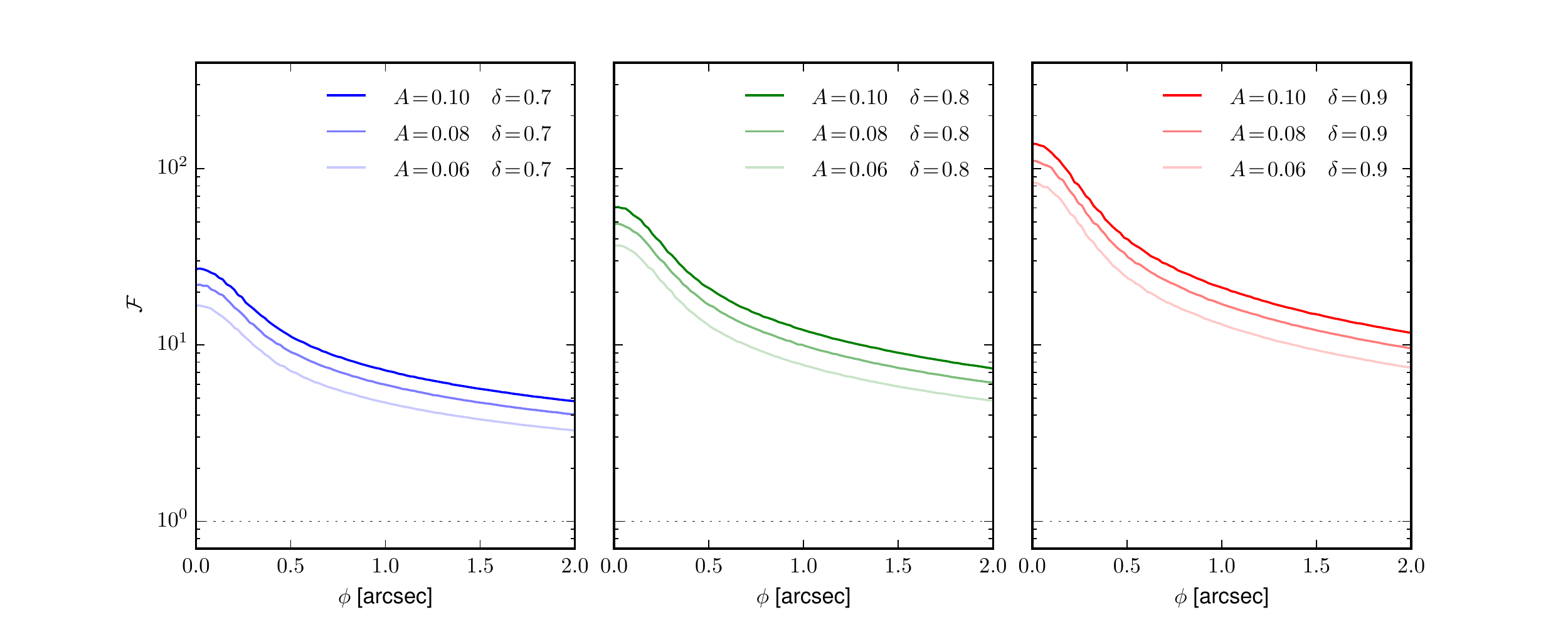}
\caption{The ratio $\cal{F}$ compares the Bayes factors with and without clustering as a function of angular separation $\phi$. The different graphs illustrate the effect for a range of $A$ and $\delta$ parameters of the angular correlation function. The sources are assumed to have the same astrometric uncertainty of \mbox{$\sigma\!=\!0.1''$}. The panels from left to right show the results for \mbox{$\delta = 0.6, 0.8, 1$}, respectively. Within each panel the three curves correspond to $A = 0.1, 0.08, 0.06$ with decreasing intensity. As expected, the stronger the clustering (larger $A$ and $\delta$ values), the more significant the correction is. The dotted line at \mbox{${\cal{}F}\!=\!1$} corresponds to no correlations. }
\label{fig:correction}
\end{figure*}

\begin{figure*}
\centering
\includegraphics[trim=20mm 0mm 20mm 7mm,clip,width=\textwidth]{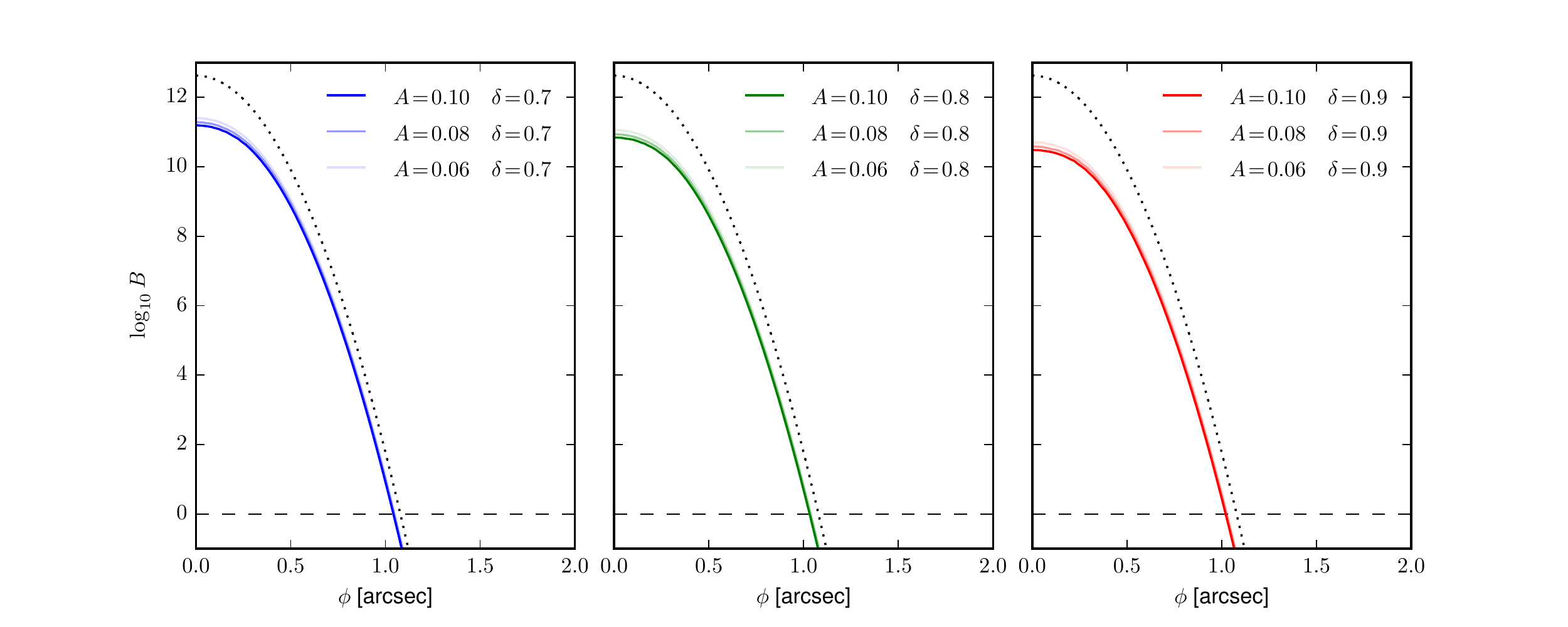}
\caption{The reduced Bayes factor as a function of angular separation. The panels correspond to those in Figure~\ref{fig:correction} and demonstrate the reduction in the evidence in the presence of clustering. While the slope $\delta$ has measurable effect across the panels, the  different amplitudes $A$ yield practically indistinguishable curves on these scales.}
\label{fig:bf}
\end{figure*}

\section{Numerical Evaluation}
\label{sec:eval}
The updated Bayes factor and its denominator can be evaluated using Monte-Carlo integration but we focus our attention directly to the correction factor. The ratio of the two Bayes factors is a function of all relevant parameters: 
\begin{equation}
\mathcal{F}(\phi;  \sigma_1,\sigma_2, A,\delta)
= 
\frac{B(\phi; \sigma_1,\sigma_2)}{B'(\phi; \sigma_1,\sigma_2, A,\delta)} \,,
\end{equation}
where $\phi$ is the angular separation between the $\xvec_1$ and $\xvec_2$ measured directions.
Looking at the definitions in eqs.~(\ref{eqn:expanded_original_bayes})
and (\ref{eq:bf_new})
we see that the numerators cancel out in the ratio.
If we further assume an all-sky coverage and hence isotropic prior distribution, e.g., on the $\nvec_1$ variables of both models, the normalization constants of $4\pi$ cancels out, and we arrive at
\begin{equation}
\mathcal{F} = \frac{\int\!\!\int \ell_1(\nvec_1)\,\ell_2(\nvec_2)\big[1\!+\!w(\theta_{12})\big]  d\nvec_1 d\nvec_2}
{\int\!\ell_1(\nvec_1)\,d\nvec_1\ \int\!\ell_2(\nvec_2)\,d\nvec_2} \,.
\label{eqn:correction_inverse}
\end{equation}
This result actually holds for partial sky coverage as well for objects far from the edge of the field of view \citep{BS08-BayesCrossID}.
Gaussian likelihood functions $\ell_1$ and $\ell_2$ further simplify the numerical problem. 
Because eq.~(\ref{eq:multivariate-gaussian}) is symmetric in $\xvec$ and $\nvec$, and hence integrates to $1$ over $\nvec$, the approximate solution becomes 
\begin{equation}
\mathcal{F} \approx 1 + \Big\langle w(\theta_{ij}) \Big\rangle \,,  
\end{equation}
where the sample of \mbox{$\{ \theta_{ij}\}$} angular separations are obtained from random pairs of ($\nvec_1$,$\nvec_2$) of unit vectors drawn from $\ell_1$ and $\ell_2$ given the observed ($\xvec_1$,$\xvec_2$) directions. The numerical averaging, noted by $\langle{}\dots\rangle$, is over this sample of angles.

Figure~\ref{fig:correction} illustrates this correction in a scenario consistent with the astrometric accuracy of the Sloan Digital Sky Survey \citep{sdss-astrometry} with \mbox{$\sigma_1\!=\!\sigma_2\!=\!0.1''$}.
The correction factor was obtained for each observed separation $\phi$ by averaging 4 million random pairs of directions ($\nvec_1$, $\nvec_2$) for high accuracy. 
From left to right the panels show 
the strong effect of increasing the slope $\delta$ from $0.6$ to $1$.
Within each plot, however, the curves 
for various amplitudes $A$ in the interval $[0.06, 0.1]$ differ only slightly.

%
By applying these corrections to the analytic results of \citet{BS08-BayesCrossID}
\begin{equation}
B(\phi;\sigma_1,\sigma_2) = \frac{2}{\sigma_1^2\!+\!\sigma_2^2}\,\exp\!\left[-\frac{\phi^2}{2(\sigma_1^2\!+\!\sigma_2^2)}\right] \,,
\label{eqn:analytic_bayes}
\end{equation}
we can see the effects on the Bayes factors in Figure~\ref{fig:bf}. The panels lines and colors are organized identically to that in Figure~\ref{fig:correction} for easy comparison. We see that at these scales the effect of the amplitude $A$ is practically indistinguishable but the slope of the correlation function $\delta$ has a big impact. 
The dashed line at zero corresponds to the boundary above which the data favor a match. Note that eq.~(\ref{eqn:analytic_bayes}) does not hold if there are other sources nearby with a non-negligible Bayes factor.

\begin{figure*}
\centering
\includegraphics[trim=20mm 10mm 20mm 15mm,clip,width=0.9\textwidth]{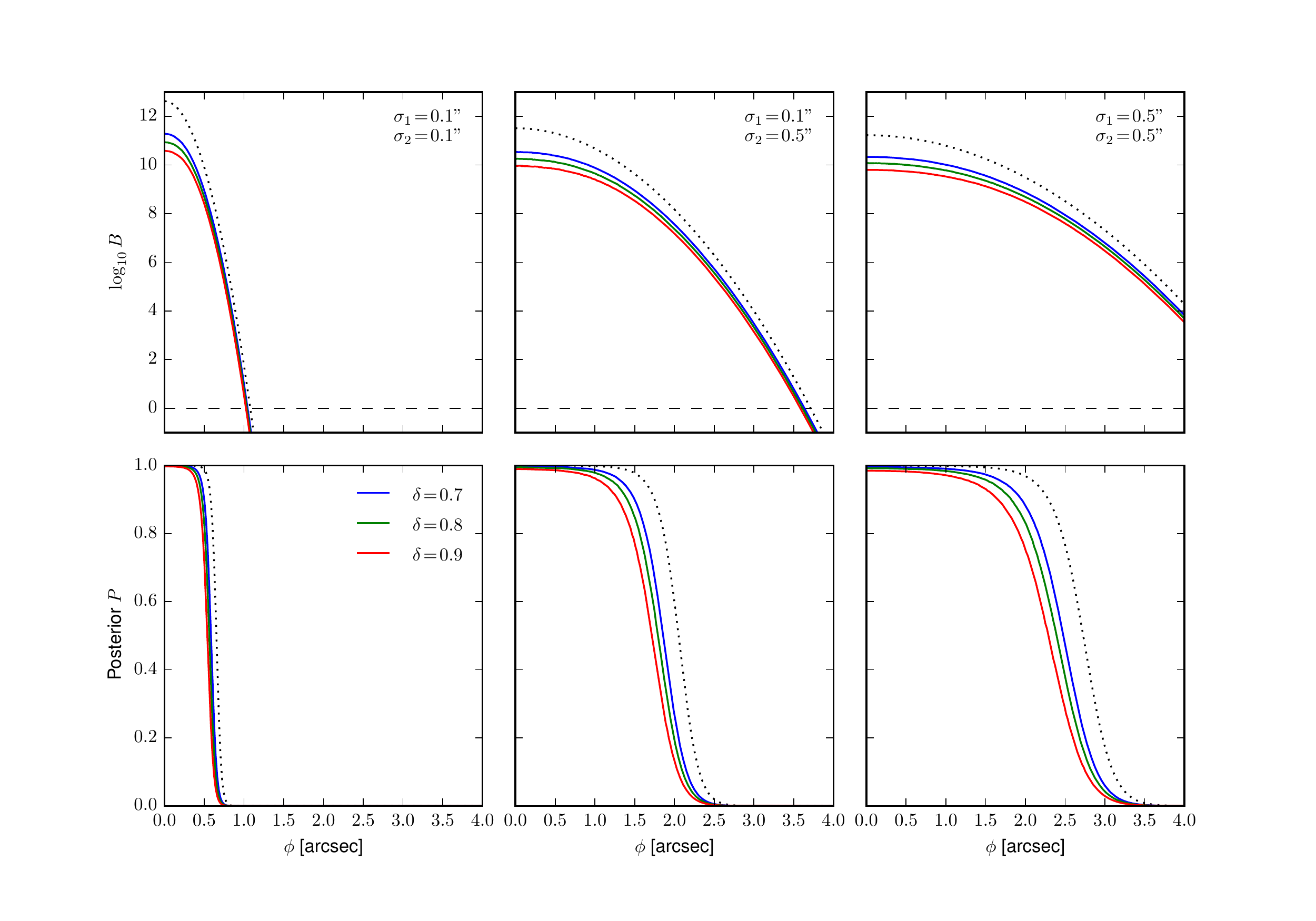}
\caption{The corrected Bayes factors (top panels) and posterior probabilities (bottom panels) are shown as a function of angular separation of the true object directions in three scenarios of matching detections with $0.1''$ and $0.5''$ astrometric uncertainties (see text).}
\label{fig:prob}
\end{figure*}

Strictly speaking, \mbox{$B\!=\!1$} really corresponds to the special case when the posterior probability $P$ is the same as the prior $P_0$  of the hypothesis that assumes a true association. In general the dependence is given by
\begin{equation}
P = \left[1+\frac{1\!-\!P_0}{B P_0}\right]^{-1} .
\end{equation}
The prior $P_0$ is primarily a function of the average surface density of sources in the catalogs and of the overlap of the two selections functions, which can be accurately taken into account using maximum likelihood estimates \citep{budavari_loredo}. 
For illustration purposes we pick \mbox{$P_0\!=\!10^{-8}$}, which corresponds to a density of about 1.5 galaxies per square arc minute.  Figure~\ref{fig:prob} shows the resulting Bayes factors (top panels) and the posterior probabilities (bottom panels) in three different matching scenarios with $0.1''$ and $0.5''$ astrometric uncertainties. For these curves we opted for the fiducial \mbox{$A\!=\!0.08$} amplitude. 
In essence the top left panel is a summary of Figure~\ref{fig:bf} and the corresponding probability curves are just below it. We see that due to the clustering a drop in the posterior occurs at shorter separations.
The middle and right panels illustrate the same trends for different $\sigma_1$ and $\sigma_2$ accuracies. The difference is more visible in these scenarios.

\section{Discussion}
\label{sec:sum}
The new probability model for the cross-identification of galaxies 
introduced in this paper enables us to study the effect of clustering strength on catalog matching. 
While at the shortest separations the effect on the Bayes factor can be as large as a factor of a hundred, and the posterior of each match becomes lower, the transition scale in the probability as a function of the angular distance is only slightly shifted.
We note that due to the consistently lower probabilities, the original matching procedure (using analytic evidence calculation) can be thought of as a fast pre-filter for catalogs, since that selection criterion is more inclusive. In other words, in practice one can use the independent model to look for candidates but use the corrected values for resolving multiple matches in crowded regions if necessary.

In our experiments we found that while the corrected $\log_{10}B'$ was not a simple quadratic function of the separation, it was always possible to accurately approximate the relevant part of the peak with even polynomials of $\phi$, i.e.,
\begin{equation}
\log_{10}B'(\phi) \approx \sum_{k=0}^K \beta_k\phi^{2k} \,.
\end{equation}
The (generalized) linear fitting can be easily solved for the $\boldsymbol{\beta}$ coefficients. We found that only 4-5 coefficients would produce fits without systematics left in the residuals.
Such approximation can be used in a practical setting for high-speed evaluation of the corrected Bayes factor even if interpolation is cumbersome, e.g., in a data\-base SQL query.

The one caveat of the analysis is that the angular correlation function is not very well known at the smallest separations. 
In principle, changes to that end of the correlation function can modulate the conditional prior $p(\nvec_2|\nvec_1)$ and the integral going from $\theta=0$. At such short scales, however, the volume of the integral shrinks very fast and hence the effect would be relatively small.
Finally, the physical size of the galaxies should eventually put an end to the increase in the real space clustering. These effects are expected to make the correction factors smaller.

\section*{Acknowledgments}

TB acknowledges partial support from NSF Grant AST-1412566 and NASA via the awards NNG16PJ23C and STScI-49721 under NAS5-26555. 

\section*{References}

\newcommand{\apj}{Astrophysical Journal}
\newcommand{\apjl}{Astrophysical Journal Letters}
\newcommand{\aj}{Astronomical Journal}
\newcommand{\mnras}{Monthly Notices of the Royal Astronomical Society}
\newcommand{\jcap}{Journal of Cosmology and Astroparticle Physics}
\newcommand{\aap}{Astronomy \& Astrophysics}
\newcommand{\aaps}{Astronomy \& Astrophysics, Supplement}
\newcommand{\prd}{Physical Review D}
\newcommand{\pasp}{Publications of the Astronomical Society of the Pacific}

\bibliographystyle{plainnat} 

\setcitestyle{authoryear,open={((},close={))}}
 
\bibliography{xmatch2.bib}

\end{document}